\documentclass[english,aps,manuscript]{revtex4}
\usepackage[T1]{fontenc}
\usepackage[latin9]{inputenc}
\usepackage{textcomp}
\usepackage{amsmath}
\usepackage{amssymb}
\usepackage{graphicx}

\makeatletter
\@ifundefined{textcolor}{}
{%
 \definecolor{BLACK}{gray}{0}
 \definecolor{WHITE}{gray}{1}
 \definecolor{RED}{rgb}{1,0,0}
 \definecolor{GREEN}{rgb}{0,1,0}
 \definecolor{BLUE}{rgb}{0,0,1}
 \definecolor{CYAN}{cmyk}{1,0,0,0}
 \definecolor{MAGENTA}{cmyk}{0,1,0,0}
 \definecolor{YELLOW}{cmyk}{0,0,1,0}
 }

\makeatother

\usepackage{babel}
\begin{document}

\title{Optoelectronic Reservoir Computing}

\author{Y. Paquot$^{\text{1}*}$, F. Duport$^{\text{1*}}$, A. Smerieri$^{\text{1}*}$,
J. Dambre$^{\text{2}}$, \\
B. Schrauwen$^{\text{2}}$, M. Haelterman$^{\text{1}}$, S. Massar$^{3\dagger}$}

\address{$^{\text{1}}$Service OPERA-photonique, Universit� libre de Bruxelles
(U.L.B.), 50 Avenue F. D. Roosevelt, CP 194/5, B-1050 Bruxelles, Belgium}

\address{$^{\text{2}}$Department of Electronics and Information Systems (ELIS),
Ghent University, Sint-Pietersnieuwstraat 41, 9000 Ghent, Belgium.}

\address{$^{\text{3}}$Laboratoire d'Information Quantique (LIQ), Universit�
libre de Bruxelles (U.L.B.), 50 Avenue F. D. Roosevelt, CP 225, B-1050
Bruxelles, Belgium}

\address{$^{\text{*}}$These authors contributed equally to this work}

\address{$^{\dagger}$ Corresponding author: smassar@ulb.ac.be}
\begin{abstract}
Reservoir computing is a recently introduced, highly efficient bio-inspired
approach for processing time dependent data. The basic scheme of reservoir
computing consists of a non linear recurrent dynamical system coupled
to a single input layer and a single output layer. Within these constraints
many implementations are possible. Here we report an opto-electronic
implementation of reservoir computing based on a recently proposed
architecture consisting of a single non linear node and a delay line.
Our implementation is sufficiently fast for real time information
processing. We illustrate its performance on tasks of practical importance
such as nonlinear channel equalization and speech recognition, and
obtain results comparable to state of the art digital implementations. 
\end{abstract}
\maketitle
\pagebreak{}

\section{Introduction}

The remarkable speed and multiplexing capability of optics makes it
very attractive for information processing. These features have enabled
the telecommunications revolution of the past decades. However, so
far they have not been exploited insomuch as computation is concerned.
The reason is that optical nonlinearities are very difficult to harness:
it remains challenging to just demonstrate optical logic gates, let
alone compete with digital electronics\cite{Caulfield2010}. This
suggests that a much more flexible approach is called for, which would
exploit as much as possible the strenghts of optics without trying
to mimic digital electronics. Reservoir computing \cite{Jaeger2001,Jaeger2001a,Jaeger2004,Legenstein2005,Maass2002,Steil,Verstraeten2007,Lukosevicius2009,Hammer2009},
a recently introduced, bio-inspired approach to artificial intelligence,
may provide such an opportunity. 

Here we report the first experimental reservoir computer based on
an opto-electronic architecture. As nonlinear element we exploit the
sine nonlinearity of an integrated Mach-Zehnder intensity modulator
(a well known, off-the-shelf component in the telecommunications industry),
and to store the internal states of the reservoir computer we use
a fiber optics spool. We report results comparable to state of the
art digital implementations for two tasks of practical importance:
nonlinear channel equalization and speech recognition.

Reservoir computing, which is at the heart of the present work, is
a highly successful method for processing time dependent information.
It provides state of the art performance for tasks such as time series
prediction \cite{Jaeger2004} (and notably won a financial time series
prediction competition\cite{NeuralForecastingCompetition}), nonlinear
channel equalization \cite{Jaeger2004}, or speech recognition \cite{Verstraeten2006,Triefenbach2010,Jaeger2007}.
For some of these tasks reservoir computing is in fact the most powerful
approach known at present. 

The central part of a reservoir computer is a nonlinear recurrent
dynamical system that is driven by one or multiple input signals.
The key insight behind reservoir computing is that the reservoir's
response to the input signal, i.e., the way the internal variables
depend on present and past inputs, is a form of computation. Experience
shows that in many cases the computation carried out by reservoirs,
even randomly chosen ones, can be extremely powerful. The reservoir
should have a large number of internal (state) variables. The exact
structure of the reservoir is not essential: for instance, in some
works the reservoir closely mimics the interconnections and dynamics
of biological neurons in a brain \cite{Maass2002}, but many other
architectures are possible. 

To achieve useful computation on time dependent input signals, a good
reservoir should be able to compute a large number of different functions
of its inputs. That is, the reservoir should be sufficiently high-dimensional,
and its responses should not only depend on present inputs but also
on inputs up to some finite time in the past. To achieve this, the
reservoir must operate at the threshold of dynamical instability which
ensures that the system exhibits a {}``fading memory'', meaning
that it will gradually forget previous inputs.

Reservoir computing is a versatile and flexible concept. This follows
from two key points: 1) many of the details of the nonlinear reservoir
itself are unimportant except for the dynamic regime (threshold of
instability) which can be tuned by some global parameters; and 2)
the only part of the system that is trained is a linear output layer.
Because of this flexibility, reservoir computing is amenable to a
large number of experimental implementations. Thus proof of principle
demonstrations have been realized in a bucket of water \cite{Fernando2003}
and using an analog VLSI chip \cite{Schurmann2004}, and arrays of
semiconductor amplifiers have been considered in simulation \cite{Vandoorne2008}.
However, it is only very recently that an analog implementation with
performance comparable to digital implementations has been reported:
namely, the electronic implementation presented in \cite{Appeltant}.

Our experiment is based on a similar architecture as that of \cite{Appeltant},
namely a single non linear node and a delay line. The main differences
are the type of non linearity and the desynchronisation of the input
with respect to the period of the delay line. These differences highlight
the flexibility of the concept. The performance of our experiment
on two benchmark tasks, isolated digit recognition and non linear
channel equalization, is comparable to state of the art digital implementations
of reservoir computing. Compared to\cite{Appeltant}, our experiment
is almost 6 orders of magnitude faster, and a further 2-3 orders of
magnitude speed increase should be possible with only small changes
to the system. 

The flexibility of reservoir computing and its success on hard classification
tasks makes it a promising route for realizing computation in physical
systems other than digital electronics. In particular it may provide
innovative solutions for ultra fast or ultra low power computation.
In Supplementary Material 1 we describe reservoir computing in more
detail and provide a road map for building high performance analog
reservoir computers. In the discussion we discuss how to measure the
speed of experimental reservoir computers.

\section{Results}

\subsection{Principles of Reservoir Computing}

Before introducing our implementation, we recall a few key features
of reservoir computing; for a more detailed treatment of the underlying
theory, we refer the reader to Supplementary Material 1. 

As is traditional in the literature, we will consider tasks that are
defined in discrete time, e.g., using sampled signals. We denote by
$u(n)$ the input signal, where $n\in$$\mathbb{Z}$ is the discretized
time; by $\overline{x}(n)$ the internal states of the system used
as reservoir; and by $\hat{y}(n)$ the output of the reservoir. A
typical evoultion law for $\overline{x}(n)$ is $\overline{x}(n+1)=f(\mathbf{A}\overline{x}(n)+\overline{m}u(n))$,
where $f$ is a nonlinear function, $\mathbf{A}$ is the time independent
connection matrix and $\overline{m}$ is the time independent input
mask. Note that in our work we will use a slightly different form
for the evolution law, as explained below. 

In order to perform the computation one needs a readout mechanism.
To this end we define a subset $x_{i}(n),\:0\leq i\leq N-1$ (also
in discrete time) of the internal states of the reservoir. It is these
states which are observed and used to build the output. The time dependent
output is obtained in an output layer by taking a linear combination
of the internal states of the reservoir $\hat{y}(n)=\sum_{i=0}^{N-1}W_{i}x_{i}(n)$.
The readout weights $W_{i}$ are chosen to minimize the Mean Square
Error (MSE) between the estimator $\hat{y}(n)$ and a target function
$y(n)$: 
\begin{equation}
MSE=\frac{1}{L}\sum_{n=1}^{L}\left(y(n)-\hat{y}(n)\right)^{2}\label{eq:MSEdefinition}
\end{equation}
over a set of examples (the training set). Because the MSE is a quadratic
function of the $W_{i}$ the optimal weights can be easily computed
from the knowledge of $x_{i}(n)$ and $y(n)$. In a typical run, the
quality of the reservoir is then evaluated on a second set of examples
(the test set). After training, the $W_{i}$ are kept fixed.

\subsection{Principles of our implementation}

In the present work we use an architecture related to that used in
\cite{Appeltant} and to the minimum complexity networks studied in
\cite{Rodan2011}. As in \cite{Appeltant}, the reservoir is based
on a non-linear system with delayed feedback (a class of systems widely
studied in the nonlinear dynamics community, see e.g. \cite{Erneux})
and consists of a single nonlinear node and a delay loop. The information
about the previous internal state of the reservoir up to some time
$T$ in the past is stored in the delay loop. After a period $T$
of the loop, the entire internal state has been updated (processed)
by the nonlinear node. In contrast to the work described in \cite{Appeltant},
the nonlinear node in our implementation is essentially instantaneous.
Hence, in the absence of input, the dynamics of our system can be
approximated by the simple recursion 
\begin{equation}
x(t)=\sin\left(\alpha\cdot x\left(t-T\right)+\varphi\right)\label{eq:SimplifiedDynamics}
\end{equation}
where $\alpha$ (the \emph{feedback gain}) and $\varphi$ (the \emph{bias})
are adjustable parameters and we have explicitly written the sine
nonlinearity used in our implementation.

We will use this system to perform useful computation on input signals
$u(n)$ evolving in discrete time $n\in\mathbb{Z}.$ As the system
itself operates in continuous time, we need to define ways to convert
input signal(s) to continuous time and to convert the system's state
back to discrete time. The first is achieved by using a sample and
hold procedure. We obtain a piecewise constant function $u(t)$ of
the continuous variable $t$ : $u(t)=u(n),\: nT'\leq t<(n+1)T'$.
The time $T'\leq T$ is taken to be less than or equal to the period
$T$ of the delay loop; when $T'\neq T$ we are in the unsynchronised
regime (see below). To discretize the system's state, we note that
the delay line acts as a memory, storing the delayed states of the
nonlinearity. From this large-dimensional state space, we take $N$
samples by dividing the input period $T'$ into $N$ segments, each
of duration $\theta$ and sampling the state of the delay line at
a single point with periodicity $\theta$. This provides us with $N$
\emph{snapshots} of the nonlinearity's reponse to each input sample
$u(n)$. From these snapshots, we construct $N$ discrete-time sequences
$x_{i}(n)=x(nT'+(i+1/2)\theta)$ ($i=0,1,...N-1$) to be used as reservoir
states from which the required (discrete-time) output is to be constructed.

Without further measures, all such recorded reservoir states would
be identical, so for computational purposes our system is one-dimensional.
In order to use this system as a reservoir computer, we need to drive
it in such a way that the $x_{i}(n)$ represent a rich variety of
functions of the input history. It is often helpful \cite{Lukosevicius2009,Rodan2011}
to use an {}``input mask'' that breaks the symmetry of the system.
In \cite{Appeltant} good performance was improved by using a nonlinear
node with an intrinsic time scale longer than the time scale of the
input mask. In the present work we also use the {}``input mask'',
but as our nonlinearity is instantaneous, we cannot exploit its intrinsic
time scale. We instead chose to desynchronize the input and the reservoir,
that is, we hold the input for a time $T'$ which differs slightly
from the period $T$ of the delay loop. This allows us to use each
reservoir state at time $n$ for the generation of a new different
state at time $n+1$ (unlike the solution used in \cite{Appeltant}
where the intrinsic time scale of the nonlinear node makes the successive
states highly correlated). We now explain these important notions
in more detail. 

The input mask $m(t)=m(t+T')$ is a periodic function of period $T'$.
It is piecewise constant over intervals of length $\theta$, i.e.,
$m(t)=m_{j}$ when $nT'+j\theta\leq t<nT'+(j+1)\theta$, for $j=0,1,...,N-1$.
The values $m_{j}$ of the mask are randomly chosen from some probability
distribution. The reservoir is driven by the product $v(t)=\beta m(t)u(t)$
of the input and the mask, with $\beta$ an adjustable parameter (the
\emph{input gain}). The dynamics of the driven system can thus be
approximated by 
\begin{equation}
x(t)=\sin\left(\alpha x(t-T)+\beta m(t)u(t)+\varphi\right)\label{eq:dynamics}
\end{equation}
It follows that the reservoir states can be approximated by 
\begin{equation}
x_{i}(n)=sin\left(\alpha x_{i}(n-1)+\beta m_{i}u(n)+\varphi\right)\label{eq:discretedynamics}
\end{equation}
when $T'=T$ (the synchronized regime) or more generally as 
\begin{equation}
x_{i}\left(n\right)=\begin{cases}
\sin\left(\alpha x_{i-k}\left(n-1\right)+\beta m_{i}u\left(n\right)+\varphi\right) & k\leq i<N\\
\sin\left(\alpha x_{N+i-k}\left(n-2\right)+\beta m_{i}u\left(n\right)+\varphi\right) & 0\leq i<k
\end{cases}\label{eq:discreteunsynchronizeddynamics}
\end{equation}
when $T'=\frac{N}{N+k}T$, ($k\in\{1,...,N-1\}$) (the unsynchronized
regime). In the synchronized regime, the reservoir states correspond
to the responses of $N$ uncoupled discrete-time dynamical systems
which are similar, but slightly different through the randomly chosen
$m_{j}$. In the unsynchronized regime, with a desynchronization $T-T'=k\theta$
, the state equations become coupled, yielding a much richer dynamics.
With an instantaneous nonlinearity, desynchronisation is necessary
to obtain a set of state transformations that is useful for reservoir
computing. We beleive that it will also be useful when the non linearity
has an intrinsic time scale, as it provides a very simple way to enrich
the dynamics.

In summary, by using an input mask, combined with desynchronization
of the input and the feedback delay, we have turned a system with
a one-dimensional information representation into an N-dimensional
system.

\subsection{Hardware setup}

\begin{figure}[ph]
\includegraphics[scale=0.5]{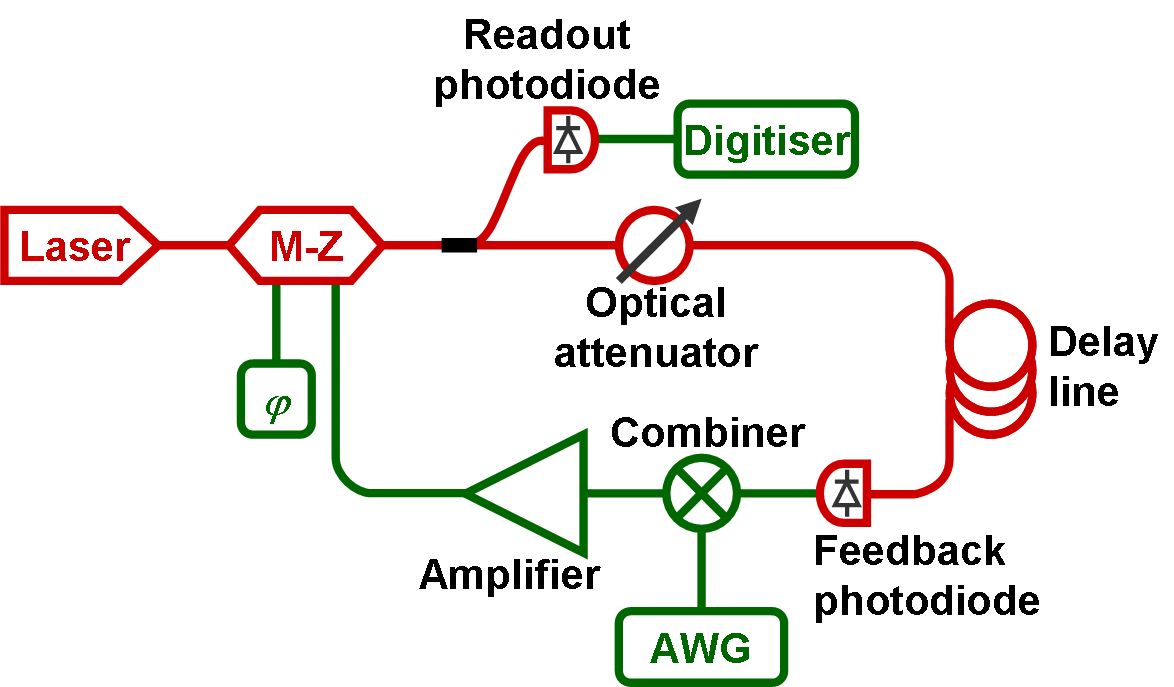}

\caption{\label{fig:Experimental-set-up.}Scheme for the experimental set-up.
The red and green parts represent respectively the optical and electronic
components. The optical part of the setup is fiber based, and operates
around 1550nm (standard telecommunication wavelength). {}``M-Z''
: Lithium Niobate Mach-Zehnder modulator. {}``$\varphi$'': DC voltage
determining the operating point of the M-Z modulator. {}``Combiner''
: electronic coupler adding the feedback and input signals. {}``AWG'':
arbitrary waveform generator. A computer generates the input signal
for a task and feeds it into the system using the arbitrary waveform
generator. The response of the system is recorded by a digitiser and
retrieved by the computer which optimizes the read-out function in
a post processing stage. The feedback gain $\alpha$ is adjusted by
changing the average intensity inside the loop with the optical attenuator.
The input gain $\beta$ is adjusted by changing the output voltage
of the function generator by a multiplicative factor. The bias $\varphi$
is adjusted by using a DC voltage to change the operating point of
the M-Z modulator. The operation of the system is fully automated
and controlled by a computer using MATLAB scripts.}
\end{figure}

\begin{figure}[h]
\includegraphics[scale=0.4]{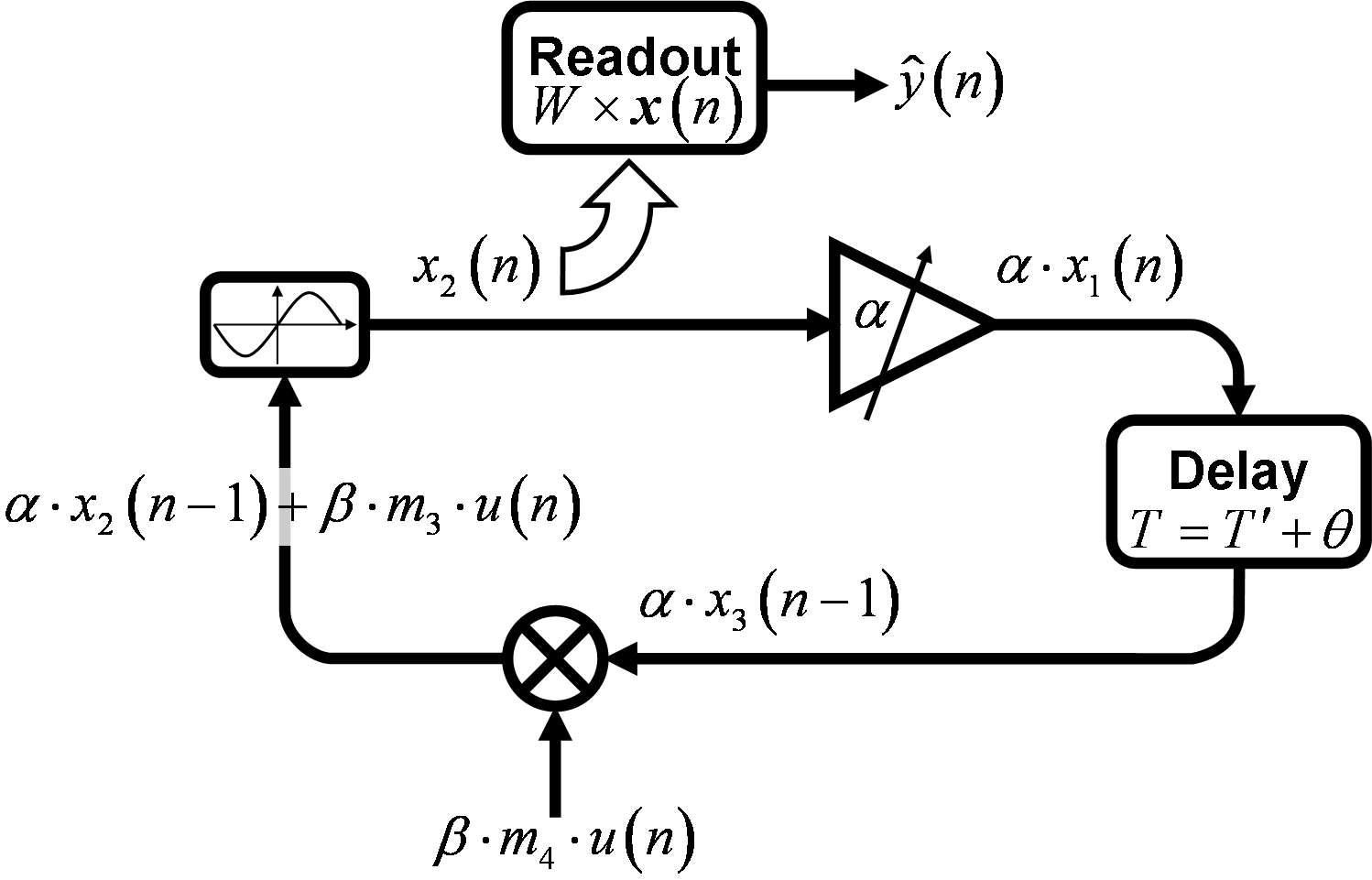}\caption{\label{fig:diagram}Schematic diagram of the information flow in the
experiment depicted in Fig. \ref{fig:Experimental-set-up.}. On the
plot we have represented four reservoir nodes at different stages
of processing, labelled according to equation \ref{eq:discreteunsynchronizeddynamics}
with $k=1.$ Starting from the bottom, and going clockwise, a input
value $u(n)$ gets multiplied by an input gain $\beta$ and a mask
value $m_{i}$, then mixed with the previous node state $\alpha x_{i-k}(n-1)$.
The result goes through the sine function to give the new state of
the reservoir $x_{i}(n)$, which then gets amplified by a factor $\alpha$
and, after the delay, will get mixed with a new input $u(n+1)$. All
the network states $x_{i}(n)$ are also collected by the readout unit,
multiplied by their respective weights $W_{i}$ and added together
to give the desired output $\hat{y}(n)$.}

\end{figure}

The above architecture is implemented in the experiment depicted in
Fig. \ref{fig:Experimental-set-up.}. The sine nonlinearity is implemented
by a voltage driven intensity modulator (Lithium Niobate Mach Zehnder
interferometer), placed at the output of a continuous light source,
and the delay loop is a fiber spool. A photodiode converts the light
intensity $I(t)$ at the output of the fiber spool into a voltage;
this is mixed with an input voltage generated by a function generator
and proportional to $m(t)u(t)$, amplified, and then used to drive
the intensity modulator. The feedback gain $\alpha$ is set by adjusting
the average intensity $I_{0}$ of the light inside the fiber loop
with an optical attenuator. By changing $\alpha$ we can bring the
system to the edge of dynamical instability. The nonlinear dynamics
of this system have already been extensively studied, see \cite{Larger2005,ChemboKouomou2005,Peil2009}.
The dynamical variable $x(t)$ is obtained by rescaling the light
intensity to lie in the interval $[-1,+1]$ through $x(t)=2I(t)/I_{0}-1$.
Then, neglecting the effect of the bandpass filter induced by the
electronic amplifiers, the dynamics of the system is given by eq.
(\ref{eq:dynamics}) where $\alpha$ is proportional to $I_{0}$.
Equation (\ref{eq:dynamics}), as well as the discretized versions
thereof, eqs. (\ref{eq:discretedynamics}) and (\ref{eq:discreteunsynchronizeddynamics}),
are derived in the supplementary material; the various stages of processing
of the reservoir nodes and inputs are shown in Fig. \ref{fig:diagram}. 

In our experiment the round trip time is $T=8.504$\textmu{}s and
we typically use $N=50$ internal nodes. The parameters $\alpha$
and $\beta$ in eq. (\ref{eq:dynamics}) are adjusted for optimal
performance (their optimal value may depend on the task), while $\varphi$
is usually set to 0. The intensity $I(t)$ is recorded by a digitizer,
and the estimator $\hat{y}(n)$ is reconstructed offline on a computer.

\begin{figure}[ph]
\includegraphics[scale=0.6]{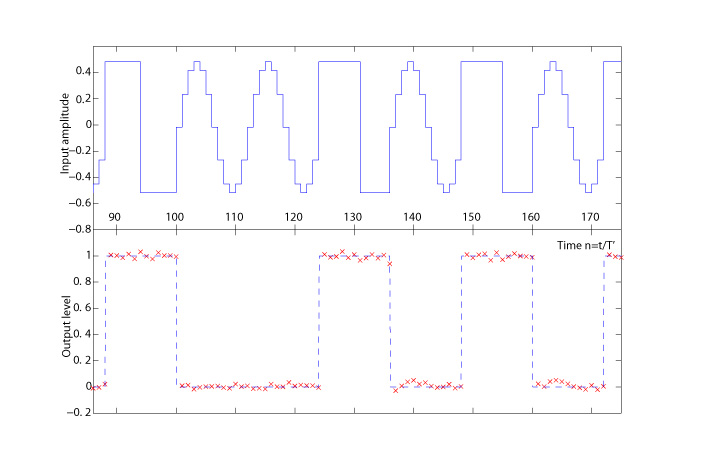}

\caption{Signal classification task. The aim is to differentiate between square
and sine waves. The top panel shows the input $u\left(t\right)$,
a stepwise constant function resulting from the discretization of
successive step and sine functions. The bottom panel shows in red
crosses the output of the reservoir $\hat{y}\left(n\right)$. The
target function (dashed line in the lower panel) is equal to 1 when
the input signal is a step function and to 0 when the input signal
is a sine function. The Normalized Mean Square Error, evaluated over
$1000$ inputs, is $\textrm{NMSE}\simeq1.5\ 10^{-3}$. \label{fig:SignalClass}}
\end{figure}

We illustrate the operations of our reservoir computer in Fig.\ref{fig:SignalClass}
where we consider a very simple signal recognition task. Here, the
input to the system is taken to be a random concatenation of sine
and square waves; the target function $y(n)$ is $0$ for a sine wave
and $1$ for a square wave. The top panel of Fig. \ref{fig:SignalClass}
shows the input to the reservoir: the blue line is the representation
of the input in continuous time $u(t)$. In the bottom panel, the
output of the network after training is shown with red crosses, against
the desired output represented by a blue line. The performance on
this task is essentially perfect: the Normalized Mean Square Error
$NMSE=\frac{1}{L}\sum_{n=1}^{L}\left(y(n)-\hat{y}(n)\right)^{2}/\frac{1}{L}\sum_{n=1}^{L}\left(y(n)\right)^{2}$%
\footnote{Note that, although reservoirs are usually trained using linear regression,
i.e., minimizing the MSE, they are often evaluated using other error
metrics. In order to be able to compare with previously reported results,
we have adopted the most commonly used error metric for each task.%
} reaches $\textrm{NMSE}\simeq1.5\cdot10^{-3}$, which is significantly
better than the results reported using simulations in \cite{Vandoorne2008}.

\subsection{Experimental results}

We have checked the performance of this system extensively in simulations.
First of all, if we neglect the effects of the bandpass filters, and
neglect all noise introduced in our experiment, we obtain a discretized
system described by eq. (\ref{eq:discreteunsynchronizeddynamics})
which is similar to (but nevertheless distinct from) the minimum complexity
reservoirs introduced in \cite{Rodan2011}. We have checked that this
discretized version of our system has performance similar to usual
reservoirs on several tasks. This shows that the chosen architecture
is capable of state of the art reservoir computing, and sets for our
experimental system a performance goal. Secondly we have also developed
a simulation code that takes into account all the noises of the experimental
components, as well as the effects of the bandpass filters. These
simulations are in very good agreement with the experimentally measured
dynamics of the system. They allow us to efficiently explore the experimental
parameter space, and to validate the experimental results. Further
details on these two simulation models are given in the supplementary
information.

We apply our optoelectronic reservoir to three tasks. These tasks
are benchmarks which have been widely used in the reservoir computing
community to evaluate the performance of reservoirs. They therefore
allow comparison between our experiment and state of the art digital
implementations of reservoir computing. 

For the first task, we train our reservoir computer to behave like
a Nonlinear Auto Regressive Moving Average equation of order 10, driven
by white noise (NARMA10). More precisely, given the white noise $u(n)$,
the reservoir should produce an output $\hat{y}(n)$ which should
be as close as possible to the response $y(n)$ of the NARMA10 model
to the same white noise. The task is described in detail in the methods
section. The performance is measured by the Normalised Mean Square
Error (NMSE) between output $\hat{y}(n)$ and target $y(n)$. For
a network of 50 nodes, both in simulations and experiment, we obtain
a $NMSE=0.168\pm0.015$. This is similar to the value obtained using
digital reservoirs of the same size. For instance a NMSE value of
$0.15\pm0.01$ is reported in \cite{Rodan2010} also for a reservoir
of size 50.

For our second task we consider a problem of practical relevance:
the equalization of a nonlinear channel. We consider a model of a
wireless communication channel in which the input signal $d(n)$ travels
through multiple paths to a nonlinear and noisy receiver. The task
is to reconstruct the input $d(n)$ from the output $u(n)$ of the
receiver. The model we use was introduced in \cite{Mathews1994} and
studied in the context of reservoir computing in \cite{Jaeger2004}.
Our results, given in Fig. \ref{fig:ChannelEqualization}, are one
order of magnitude better than those obtained in \cite{Mathews1994}
with a nonlinear adaptive filter, and comparable to those obtained
in \cite{Jaeger2004} with a digital reservoir. At 28 dB of signal
to noise ratio, for example, we obtain an error rate of $1.3\cdot10^{-4}$,
while the best error rate obtained in \cite{Mathews1994} is $4\cdot10^{-3}$
and in \cite{Jaeger2004} error rates between $10^{-4}$ and $10^{-5}$
are reported.

Finally we apply our reservoir to isolated spoken digits recognition
using a benchmark task introduced in the reservoir computing community
in \cite{Verstraeten2005}. The performance on this task is measured
using the Word Error Rate (WER) which gives the percentage of words
that are wrongly classified. Performances reported in the literature
are a WER of $0.55\%$ using a hidden Markov model \cite{Walker2004};
WERs of $4.3\%$ \cite{Verstraeten2005}, of $0.2\%$ \cite{Verstraeten2006},
of $1.3\%$ \cite{Rodan2011} for reservoir computers of different
sizes and with different post processing of the output. The experimental
reservoir presented in \cite{Appeltant} reported a WER of $0.2\%$.
Our experiment yields a WER of $0.4\%$, using a reservoir of $200$
nodes.

Further details on these tasks are given in the methods section and
in Supplementary Material 2.

\section{Discussion}

We have reported the first demonstration of an opto-electronic reservoir
computer. Our experiment has performance comparable to state of the
art digital implementations on benchmark tasks of practical relevance
such as speech recognition and channel equalization. Our work demonstrates
the flexibility of reservoir computers that can be readily reprogrammed
for different tasks. Indeed by re-optimising the output layer (that
is, choosing new readout weights $W_{k}$), and by readjusting the
operating point of the reservoir (changing the feedback gain $\alpha$,
the input gain $\beta$, and possibly the bias $\varphi$) one can
use the same reservoir for many different tasks. Using this procedure,
our experimental reservoir computer has been used successively for
tasks such as signal classification, modeling a dynamical system (NARMA10
task), speech recognition, and nonlinear channel equalization. 

We have introduced a new feature in the architecture, as compared
to the related experiment reported in \cite{Appeltant}. Namely by
desynchronizing the input with respect to the period of the reservoir
we conserve the necessary coupling between the internal states, but
make a more efficient use of the internal states as the correlations
introduced by the low pass filter in \cite{Appeltant} are not necessary.

Our experiment is also the first implementation of reservoir computing
fast enough for real time information processing. It can be converted
into a high speed reservoir computer simply by increasing the bandwidth
of all the components (an increase of at least 2 orders of magnitude
is possible with off-the-shelf optoelectronic components). We note
that in future realisations it will be necessary to have an analog
implementation of the pre-processing of the input (digitisation and
multiplication by the input mask) and of the post-processing of the
output (multiplication by output weights), rather than the digital
pre- and post-processing used in the present work.

From the point of view of applications, the present work thus constitutes
an important step towards building ultra high speed optical reservoir
computers. To help achieve this goal, in the supplementary material
we present guidelines for building experimental reservoir computers.
Whether optical implementations can eventually compete with electronic
implementations is an open question. From the fundamental point of
view, the present work helps understanding what are the minimal requirements
for high level analog information processing.

\section*{Acknowledgments. }

We would like to thank J. Van Campenhout whose suggestions initiated
this research project. S.M. acknowledges a helpful discussion at the
beginning of this project in which Ingo Fisher. All authors would
like to thank the researchers of the Photonics@be network working
on reservoir computing for numerous discussions over the duration
of this project. The authors acknowledge financial support by Interuniversity
Attraction Poles Program (Belgian Science Policy) project Photonics@be
IAP6/10 and by the Fonds de la Recherche Scientifique FRS-FNRS.

\section*{Methods}

\subsection*{NARMA10 task. }

Auto Regressive models and Moving Average models, and their generalisation
Nonlinear Auto Regressive Moving Average Models (NARMA), are widely
used to simulate time series. The NARMA10 model is given by the recurrence
\begin{equation}
y(n+1)=0.3y(n)+0.05y(n)\left(\sum_{i=0}^{9}y(n-i)\right)+1.5u(n-9)u(n)+0.1\label{eq:NarmaDefinition}
\end{equation}
where $u(n)$ is a sequence of random inputs drawn from an uniform
distribution over the interval $[0,0.5]$. The aim is to predict the
$y(n)$ knowing the $u(n).$This task was introduced in \cite{Atiya2000}.
It has been widely used as a benchmark in the reservoir computing
community, see for instance \cite{Jaeger2002,Rodan2010,Rodan2011}

\subsection*{Nonlinear channel equalization.\emph{ }}

This task was introduced in \cite{Mathews1994}, and used in the reservoir
computing community in \cite{Jaeger2004} and \cite{Rodan2010}. The
input to the channel is an i.i.d. random sequence $d(n)$ with values
from $\{-3,-1,+1,+3\}$. The signal first goes through a linear channel,
yielding 
\begin{eqnarray}
q(n) & = & 0.08d(n+2)\text{\textendash}0.12d(n+1)+d(n)+0.18d(n\text{\textendash}1)\nonumber \\
 &  & \text{\textendash}0.1d(n\text{\textendash}2)+0.091d(n\text{\textendash}3)\text{\textendash}0.05d(n\text{\textendash}4)\label{eq:ChannelEqMixing}\\
 &  & +0.04d(n\text{\textendash}5)+0.03d(n\text{\textendash}6)+0.01d(n\text{\textendash}7)\nonumber 
\end{eqnarray}
It then goes through a noisy nonlinear channel, yielding 
\begin{equation}
u(n)=q(n)+0.036q(n)^{2}\text{\textendash}0.011q(n)^{3}+\nu(n)\label{eq:ChannelEqNonlinAndNoise}
\end{equation}
 where $\nu(n)$ is an i.i.d. Gaussian noise with zero mean adjusted
in power to yield signal-to-noise ratios ranging from 12 to 32 db.
The task is, given the output $u(n)$ of the channel, to reconstruct
the input $d(n)$. The performance on this task is measured using
the Symbol Error Rate, that is the fraction of inputs $d(n)$ that
are misclassified (Ref. \cite{Rodan2010} used another error metric
on this task).

\subsection*{Isolated spoken digit recognition. }

The data for this task is taken from the NIST TI-46 corpus \cite{TexasInstruments}.
It consists of ten spoken digits (0...9), each one recorded ten times
by five different female speakers. These 500 spoken words are sampled
at 12.5 kHz. This spoken digit recording is preprocessed using the
Lyon cochlear ear model \cite{Lyon}. The input to the reservoir $u_{j}(n)$
consists of an 86-dimensional state vector ($j=1,...,86$) with up
to 130 time steps. The number of variables is taken to be $N=200$.
The input mask is taken to be a $N\times86$ dimensional matrix $b_{ij}$
with elements taken from the the set $\{-0.1,+0.1\}$ with equal probabilities.
The product $\sum_{j}b_{ij}u_{j}(n)$ of the mask with the input is
used to drive the reservoir. Ten linear classifiers $\hat{y}_{k}(n)$
($k=0,...,9$) are trained, each one associated to one digit. The
target function for $y_{k}(n)$ is +1 if the spoken digit is $k$,
and -1 otherwise. The classifiers are averaged in time, and a winner-takes-all
approach is applied to select the actual digit.

Using a standard cross-validation procedure, the 500 spoken words
are divided in five subsets. We trained the reservoir on four of the
subsets, and then tested it on the fifth one. This is repeated five
times, each time using a different subset as test, and the average
performance is computed. The performance is given in terms of the
Word Error Rate, that is the fraction of digits that are misclassified.
We obtain a WER of $0.4\%$ (which correspond to 2 errors in 500 recognized
digits).

\clearpage{}

\begin{figure}[ph]
\includegraphics[scale=0.5]{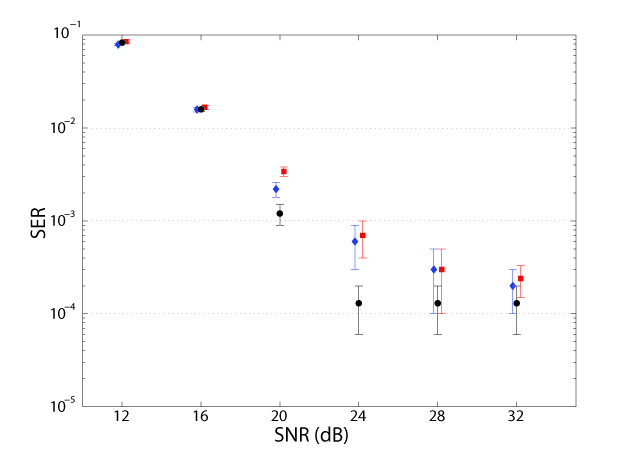}\caption{Results for nonlinear channel equalization. The horizontal axis is
the Signal to Noise Ratio (SNR) of the channel. The vertical axis
is the Symbol Error Rate (SER), that is the fraction of input symbols
that are misclassified. Results are plotted for the experimental setup
(black circles), the discrete simulations based on eq. (\ref{eq:discreteunsynchronizeddynamics})
(blue rhombs), and the continuous simulations that take into account
noise and bandpass filters in the experiment (red squares). All three
sets of results agree within the statistical error bars. Error bars
on the experimental points relative to 24, 28 and 32 dB might be overestimated
(see Supplementary Material 2). The results are practically identical
to those obtained using a digital reservoir in \cite{Jaeger2004}.
\label{fig:ChannelEqualization}}
\end{figure}
\pagebreak{}

\appendix

\part*{Supplementary Material 1: \protect \\
An Introduction to Reservoir Computing for Physicists and Engineers}

\bigskip{}

\section*{Introduction}

Understanding how physical systems can process information is a major
scientific challenge. The tremendous progress that has been accomplished
during the past century has given rise to whole new fields of science,
such as computer science and digital (silicon) electronics, or quantum
information. Information processing by classical analog systems is
a case apart because remarkable examples are found in the biological
sciences (e.g. the brain), but our understanding is still very incomplete.
Thus, we understand at the basic level many aspects of how information
is processed using biochemical reactions in cells, or how information
is processed by neurons in a brain, to take two examples. But how
these elementary processes are organized at a higher level, and what
is the large scale architecture of these systems, still escapes us.
Understanding these issues would be of great conceptual interest.
It could also have huge technological repercussions, as it could completely
change the way we build information processing machines. That this
tremendous scope for progress exists is illustrated by the approximately
6 orders of magnitude gap in energy consumption between a brain and
a present day silicon computer%
\footnote{To simulate (with a speed reduction of 2 orders of magnitude) a cat
brain that consumes roughly 1 Watt, the Blue Gene supercomputer consumes
roughly a $10^{6}$Watt. For a presentation of the simulation, see
\cite{Ananthanarayanan2009}%
}.

So far most work on information processing in analog systems has been
based on imitating biological systems. This has given rise to the
field of artificial neural networks. More abstract approaches have
also been developed, such as hidden Markov models, or support vector
machines. Reservoir computing \cite{Jaeger2001,Jaeger2001a,Maass2002,Jaeger2004,Steil,Legenstein2005,Verstraeten2007}
provides an alternative line of attack. The interest of reservoir
computing is that 1) it proposes architectures that are quite different
from those studied up to now; and 2) despite a relatively recent history
-the first papers on the topic date from 2002-, its performances are
comparable to, and sometimes even exceed, those of other approaches
to machine learning%
\footnote{Machine learning is the field of artificial intelligence concerned
with developing algorithms whose response to the input -typically
empirical data- improves with experience.%
}. A deeper understanding of reservoir computing could provide new
insights into how analog classical systems, and in particular biological
systems, process information. Additionally it could give rise to novel
approaches for implementing computation and enable new applications. 

The aim of the present text is not to duplicate existing reviews of
reservoir computing \cite{Lukosevicius2009,Hammer2009}. Rather, we
wish to present the subject from the point of view of the physicist
or engineer who wishes to build an experimental reservoir computer.
We will argue that reservoir computing provides a quite detailed,
but also quite flexible, road map towards building physical systems
that process information with architectures very different from those
used up to now.

\section*{Understanding Reservoir Computing}

To understand the potentialities of reservoir computing, it is best
to start with an example, and then examine to what extent the example
can be generalized. Finally, on the basis of this discussion, we outline
a road map for building experimental reservoir computers.

The archetypal reservoir computer is constructed as follows. It consists
of a finite number $N$ of internal variables $x_{i}(n)$ $(i=0,...,N-1)$
evolving in discrete time $n\in\mathbb{Z}$. The evolution of the
variables $x_{i}(n)$ is perturbed by an external input $u(n)$. The
evolution equations for the reservoir is:
\begin{equation}
x_{i}(n+1)=\tanh\left[\sum_{i=0}^{N-1}a_{ij}x_{j}(n)+b_{i}u(n)\right],\label{eq:ResTanh}
\end{equation}
where the (time independent) coefficients $a_{ij}$ and $b_{i}$ are
chosen independently at random from a simple distribution, for instance
Gaussian distributions with mean zero and variances $a^{2}$ and $b^{2}$.
The dynamics of this system (called the {}``reservoir'') is fixed
(i.e. the coefficients $a_{ij}$ and $b_{i}$ are fixed) . The dimensionality
$N$ of the reservoir is typically much larger than the input dimensionality.
Given proper parameters $a^{2}$ and $b^{2}$, the reservoir will
also have memory of past inputs. The use of the $\tanh$ nonlinearity
in eq. (\ref{eq:ResTanh}) is traditional, but by no means essential,
see below.

The aim of reservoir computing is to perform a computation on the
input $u(n)$. To this end one maps the internal state of the reservoir
to an output signal denoted $\hat{y}(n)$. The output should be as
close as possible to a given target function $y(n)$. An example of
this would be the detection of a stereotypical pattern in the input
sequence such as spoken digits in a stream of audio, or a certain
header in digital communication. Another example would be trying to
predict future inputs such as electrical load prediction or financial
time-series prediction.

The output $\hat{y}(n)$ of the reservoir computer at time $n$ is
a linear combination of the high dimensional internal states of the
reservoir at time $n$: 
\begin{equation}
\hat{y}(n)=\sum_{i=0}^{N-1}W_{i}x_{i}(n).\label{eq:output}
\end{equation}
The $W_{i}$ are usuallly chosen as follows. For some time interval,
$n\in[1,T]$ we simulate the reservoir such that one knows both the
inputs $u(n)$ and the target function $y(n)$. We minimize, over
this training interval, the Mean Square Error between $\hat{y}(n)$
and $y(n)$:

\begin{equation}
\textrm{MSE}=\frac{1}{T}\sum_{n=1}^{T}\left(\hat{y}(n)-y(n)\right)^{2}.\label{eq:MSE}
\end{equation}
This can be efficiently performed using standard linear regression
techniques. {[}In cases where a classification of the inputs is desired,
one may train several outputs (one for each class) and then use a
winner takes all approach, or one may try to optimize the miss-classification
rate using techniques such as logistic regression, Gaussian discriminant
analysis, regression to the indicator function or a linear support
vector machine.{]} Note that in general one should apply some form
of regularization (to reduce the model complexity) to the readout
function eq. (\ref{eq:output}). To this end one uses optimisation
techniques that prefer models with small parameters $W_{i}$. This
is accomplished by using e.g. ridge regression.

The coefficients $W_{i}$ are now fixed and the reservoir computer
is ready for use: one can input into the reservoir new inputs $u(n)$
and obtain the corresponding output $\hat{y}(n)$. Because of the
training procedure where we ensured proper generalization, the output
will continue to be close to the target function, even for new data.
It is considered good practice to check this. To this end one estimates
the performance of the reservoir in a test phase during which new
inputs are fed into the reservoir, and one compares, using the previously
chosen $W_{i}$ and an appropriate error metric, how close the output
is to the target function.

Several remarks are now in order:
\begin{enumerate}
\item For good performance, it is necessary to scale the coefficients by
a multiplicative factor 
\[
a_{ij}\to\alpha a_{ij}\quad\mbox{and}\quad b_{i}\to\beta b_{i}\,.
\]
(In other words, if the coefficients are drawn from normal distributions
with mean zero, one adjusts the variances of the normals). These parameters
are often called the feedback gain and input gain respectively. The
necessity for this scaling can be understood intuitively as follows:
if the coefficients $a_{ij}$ are too small, then the state equation
(\ref{eq:ResTanh}) is strongly damped; if the coefficients $a_{ij}$
are too large, then the system has highly complex dynamics or is even
chaotic, and its behavior is very sensitive to small changes in the
input. The coefficients $b_{i}$ are adjusted in order to have an
appropriate ratio between the contribution of the terms $\sum_{j}a_{ij}x_{j}(n)$
and the source term $b_{i}u(n)$. They also determine how non-linearly
the inputs are expanded in the reservoir states.
\item Finding the optimal values of $W_{i}$ is immediate in the case of
linear regression. Indeed inserting eq. (\ref{eq:output}) into eq.
(\ref{eq:MSE}) and extremizing with respect to $W_{i}$, one finds
that the optimal choice is
\begin{equation}
W_{i}=\sum_{j=0}^{N-1}R_{ij}^{-1}P_{j}\label{eq:OptimalCoefficients}
\end{equation}
where $R_{ij}^{-1}$ is the inverse of the correlation matrix $R_{ij}=\frac{1}{T}\sum_{n=1}^{T}x_{i}(n)x_{j}(n)$
and $P_{j}=\frac{1}{T}\sum_{n=1}^{T}x_{j}(n)y(n)$. This should be
contrasted to what would happen if one tried to optimize the coefficients
$a_{ij}$ and $b_{i}$. In this case there are many more coefficients
to optimize, and furthermore it is difficult to find the optimal value,
as one may very easily end up in local optima. The use of a linear
readout eq. (\ref{eq:output}) is thus highly advantageous.
\item Many variants on the above architecture have been investigated with
success. We can list:

\begin{enumerate}
\item Using other nonlinearities than $\tanh$, such as the biological inspired
models which approximate the way real neurons interact with each other
(spiking neuron models).
\item Using sparse connection matrices $a_{ij}$ (i.e. matrices with most
of the elements set to zero), so as to decrease the computational
resources required to simulate the evolution.
\item Using multiple inputs $u_{k}(n)$, $k=1,...,K$. In this case one
simply adds an index to the coefficients $b_{i}$. The source term
in eq. (\ref{eq:ResTanh}) thus becomes $\sum_{k}b_{ik}u_{k}(n)$.
\item Using the output $\hat{y}(n)$ as input to the reservoir. This modification
allows training the reservoir to behave as a given dynamical system
\cite{Jaeger2001,Jaeger2004}.\label{enu:Using-the-output}
\end{enumerate}
\item The above approach is extremely powerful. To take two examples:

\begin{enumerate}
\item Given a time series $u(n)$ for times $n=1,...,T$, the task is to
guess the subsequent values of the time series $u(T+1),u(T+2),...$
The time series could for instance come from the evolution of a chaotic
dynamical system, such as the Mackey-Glass system. Using the trick
\ref{enu:Using-the-output}, results which outperform all other known
approaches for predicting chaotic time series were obtained \cite{Jaeger2004}.
Reservoir computing also won an international competition on prediction
of future evolution of financial time series \cite{NeuralForecastingCompetition}.
\item Reservoir computing has been extensively applied to speech recognition.
Using well known data sets, reservoir computing outperforms other
approaches for isolated digit recognition \cite{Verstraeten2006,Jaeger2007}.
Recently reservoir computing has been applied to a more complex task:
phoneme recognition. Performances comparable to the best alternative
methods were obtained using 3 coupled reservoirs, each with $N=20000$
variables \cite{Triefenbach2010}.
\end{enumerate}
\end{enumerate}

\section*{Road map for building information processing systems}

The basic setup of reservoir computing, although typically implemented
in software, suggests many promising new avenues to implement computation
in analog dynamic systems. The theoretical requirements for reservoir
computing to be computationally universal (in the analog sense) \cite{Maass2002}
are very loose: the reservoir is required to have fading memory, to
be excitable by the input and a high dimensional readout must be possible.
Many physical systems could be conceived that adhere to these rules
and could thus potentially be turned in universal computing machines.
However turning these general ideas into a working machine in more
difficult.

If one wishes to build an experimental reservoir computer, then it
is essential to understand what are the constraints, but also the
design freedom. In this respect, a number of important lessons can
be learned from the example presented in the previous section.
\begin{enumerate}
\item The fact that the interconnection matrix $a_{ij}$ and the coefficients
$b_{i}$ in eq. (\ref{eq:ResTanh}) are chosen at random is extremely
important. It means that fine tuning of a large number of coefficients
is not necessary for a good reservoir computer. Rather almost all
interconnection matrices will give good performance.
\item The fact that the tanh in eq. (\ref{eq:ResTanh}) can be changed into
other non-linearities is also extremely important. It means that one
does not have to imitate specific non-linear behavior (such as the
specific dynamics of neurons in a biological brain), but one can use
the non-linearities that are easily accessible experimentally.
\item The fact that the only coefficients that are task specific are the
$W_{i}$ (the weights of the readout function) implies that a given
dynamical system can be used for many different information processing
tasks. It also implies that one can separate the design and analysis
of the reservoir itself from the design and analysis of the readout
function.
\end{enumerate}
On the basis of these remarks, a reservoir computer can be built out
of a dynamical system that satisfies the following constraints:
\begin{enumerate}
\item It should consist of a large number (say 50, or more, but this depends
on the task and on the specifics of the dynamic system) dynamical
variables which are coupled together by a non-linear evolution equation.
\item The evolution of the dynamical system can be perturbed by the external
input.
\item As much as possible, one should try to break all symmetries of the
system (this is the message coming from the fact that the $a_{ij}$
and $b_{i}$ are random: there is no residual structure/symmetry in
the dynamics). To this end one should privilege dynamical systems
that depend on a large number of random parameters.
\item A few global parameters must be experimentally tunable. These global
parameters are used to adjust the operating point of the system (typically
to just below the threshold of instability), and to adjust the overall
weight given to the external inputs (this corresponds to the scaling
of the coefficients $a_{ij}$ and $b_{i}$).
\item It should be possible to read-out the state of a large number (better
all) of the dynamical variables, or at least to construct the readout
function $y=\sum_{i}W_{i}x_{i}$ with adjustable weights $W_{i}$.
Note that a reservoir computer can function satisfactorily even if
only a subset of the dynamical variables are read out.
\end{enumerate}
Once these constraints are satisfied, one can proceed to test the
system (either using numerical simulations, or using the experimental
realization). In the reservoir computing literature, there exists
a series of (somewhat) standardized tests on which to evaluate the
performance of the system. Some of these tests have a theoretical
justification (e.g. linear memory capacity); others are interesting
tasks which have been often used by the community. These tasks thus
provide benchmarks with which to compare the performances of different
reservoir computers. Possible tasks to study include:
\begin{enumerate}
\item Linear memory capacity and memory function \cite{Jaeger2001a}. In
this task the target function $y(n)=u(n-k)$ is simply the input $k$
time steps in the past. Performance on this task indicates how well
the reservoir's state can be used to recover past inputs. If the linear
memory capacity is small, then the reservoir will be unable to carry
out tasks that require long memory of the input.
\item Non-linear memory capacity, see \cite{Verstraeten,Verstraeten2010}.
In this task the target function is a non-linear function of the past
inputs, for instance the product $y(n)=u(n-k)u(n-k')$. Performance
on this task measures how much non-linear processing of the input
is carried out by the reservoir.
\item Simulating specific Nonlinear Auto Regressive Moving Average (NARMA)
systems. For the NARMA tasks, the aim is to use the reservoir to simulat
the response of a NARMA system driven by a random input. Two variants
are widely used: a 10th order system (NARMA10) introduced in \cite{Atiya2000},
and the more difficult 30th order system (NARMA30) (see for instance
\cite{Schrauwen2008} for a definition).
\item Predicting the evolution of the trajectory of chaotic attractors such
as the attractor of the Mackey-Glass system \cite{Jaeger2004}.
\item Speech recognition: Several benchmarks have been published, going
from Japanese vowel recognition \cite{Jaeger2007}, to isolated spoken
digits \cite{Verstraeten2006}, to phoneme recognition \cite{Triefenbach2010}.
\end{enumerate}
This (non-exhaustive) list of benchmarks provides a natural road-map
for future experimental reservoirs. First master easy tasks such as
signal classification, or isolated digit recognition. Then go on to
moderately harder tasks such as NARMA10 or NARMA30. Finally, one can
imagine tackling hard tasks such as phoneme recognition. In all cases,
measure the linear and nonlinear memory functions, as they will give
task independent information on how the reservoir is processing information.

This road-map is appealing. Steps along it have been followed using
reservoirs based on water waves in a bucket \cite{Fernando2003},
on the cerebral cortex of a cat \cite{Nikolic}, on analog VLSI chips
\cite{Schurmann2004}, using a (numerically simulated) array of coupled
semi-conductor optical amplifiers \cite{Vandoorne2008}. However it
is only very recently that it has been possible, using an electronics
implementation and a delayed feedback loop, to demonstrate an analog
reservoir computer with performance comparable to that of digital
reservoirs on a non trivial task (in this case isolated digit recognition)
\cite{Appeltant}.

\section*{Open questions and perspectives}

The above road map leaves open a number of important questions. We
outline the most obvious:
\begin{enumerate}
\item Effect of noise. In experimental realizations of reservoir computing,
there will inevitably be noise and imperfections. How will these affect
the performance of the scheme? There has not been an exhaustive analysis
of this issue, but a few remarks are in order. One should distinguish
between noise within the reservoir itself and noise in the readout. 

\begin{enumerate}
\item Noise in the reservoir itself is deleterious. However,\emph{ }reservoir
computers can continue to work even with moderate levels of noise.
Indeed the noise can be viewed as an unwanted input, and the aim is
to perform the desired task while ignoring this second unwanted input.
A possible approach to counteract the effect of noise would simply
be to increase the size of the reservoir. But this remains to be studied
in detail.
\item Noise in the readout can be beneficial. Indeed adding noise to the
readout is a trick, equivalent to ridge regularization, often used
in numerical simulations to increase the robustness of reservoirs. 
\end{enumerate}
\item Best non-linearity. Experience with digital simulations of reservoirs
shows that sigmoidal non-linearities, such as $\tanh$, give very
good performance. However in experimental realizations, other non-linearities
may be much easier to implement. Experience suggests that the best
non-linearity depends to some extent on the task at hand. Also sub-optimal
non-linearities can presumably be compensated by increasing the size
of the reservoir.
\item Continuous time and continuous space. In numerical simulations, it
is by far easier to work with a discrete set of variables $x_{i}(n)$,
$i=0,...,N-1$ evolving in discrete time $n\in\mathbb{Z}$. But in
experimental systems, it is often much more natural to work with continuous
variables and continuous time. A theory of reservoir computing operating
with continuous variables evolving in continuous time remains to be
written, although some theoretical progress has been made along these
lines \cite{Hermans2010}.
\end{enumerate}
Finally, what is the future of experimental reservoir computing? On
the one hand the quest is fundamental: can we build analog systems
that perform high-level information processing? Will we in this way
gain new insights into how biological information processing systems
operate? On the other hand experimental reservoir computers could
ultimately have practical applications. Indeed they are based on architectures
completely different from those used in present day digital electronics.
Presumably the place to look for applications is at the frontier where
digital electronics has difficulty coping, such as information processing
at ultra high speeds, or with ultra low energy consumption, or for
tasks which are hard to code using standard programming methods.

\pagebreak{}

\part*{Supplementary Material 2 \protect \\
Experimental System, Numerical Simulations, Tasks}

\bigskip{}

Here we describe in more detail our experimental system (schematized
in Fig. 1 of the main text), the algorithm used to simulate it numerically,
and the tasks we study to validate its performance as a reservoir.

\section*{Undriven system}

We first consider the free running dynamics of our system, in the
absence of input. 

The light source consists of a Laser at the wavelength 1564nm (in
the standard C band of optical telecommunication). It produces a time
independent intensity $I_{0}$. 

The light passes through an intensity modulator (Mach-Zehnder modulator)
driven by a time dependent voltage $V\left(t\right)$. The light intensity
just after the modulator is $I\left(t\right)=I_{0}\cos^{2}\left(\pi\frac{V(t)}{2V_{\pi}}+\phi\right)$
where $V_{\pi}$ is a constant voltage (the voltage which is needed
to go from a maximum to the next minimum of light at the output of
the modulator) and $\phi$ can be adjusted by applying a DC bias voltage
to the modulator. Taking $\phi=3\pi/4+\varphi$ allows us to rewrite

\begin{equation}
I\left(t\right)=\frac{I_{0}}{2}+\frac{I_{0}}{2}\sin\left(\frac{\pi V(t)}{V_{\pi}}+\varphi\right).\label{eq:1}
\end{equation}

The light passes trough a tunable optical attenuator, enabling the
adjustment of the loop's gain through the variation of the value $I_{0}$.
It propagates through a long spool of fiber (1.7 km of single mode
optical fiber in our experiment), and is then detected by a photodiode
integrated with a transimpedance amplifier. The resulting voltage
is amplified again (by a RF amplifier) to produce the voltage $V\left(t\right)$
that drives the intensity modulator. In our experiment, the photodiode
(with its transimpedance amplifier) and RF amplifier operate in a
linear regime, hence $V\left(t\right)$ is proportional to $I\left(t-T\right)$
where $T$ is the round trip time of the oscillator. However, one
must take into account that the photodiode acts as a lowpass filter
and that the RF amplifiers act as highpass filters. Both the photodiode
and the amplifiers also add noise. Thus, approximating the filters
by first order filters, we have $\tilde{V}\left(\omega\right)=A\frac{\omega_{0}}{\omega_{0}+j\omega}\frac{j\omega}{\omega_{1}+j\omega}e^{j\omega T}\left(\tilde{I}\left(\omega\right)+\tilde{n}\left(\omega\right)\right)$
where $j$ is the imaginary unit, tilde denotes the Fourier transform,
$A$ denotes the amplification factor, $\omega_{1}\ll\omega_{0}$
are the cutoff frequencies of the resulting bandpass filter, and $n$(t)
is the white noise. In the time domain we have $V(t)=A\mathcal{F}_{\omega_{0}\omega_{1}}\left(I\left(t-T\right)+n\left(t\right)\right)$
where $\mathcal{F}_{\omega_{0}\omega_{1}}$ denotes the linear bandpass
filter. 

Because of the high pass filter, we can use as variable the fluctuations
around the mean intensity. We define $x\left(t\right)=\frac{I(t)-I_{0}/2}{I_{0}/2}$
(i.e., we rescale the intensity I(t) to lie in the interval $[-1,+1]$)
whereupon eq. (\ref{eq:1}) becomes 
\begin{equation}
x\left(t\right)=\sin\left(\alpha\mathcal{F}_{\omega_{0}\omega_{1}}\left(x\left(t-T\right)+\frac{n\left(t\right)}{I_{0}}\right)+\varphi\right)\label{eq:xsinx}
\end{equation}
where $\alpha=\frac{\pi AI_{0}}{V_{\pi}}$ is experimentally adjustable
by changing the input light intensity $I_{0}$. Experimentally it
can be tuned over the range $\alpha\in[0,4.2]$.

Typical values of the parameters are $T=8.5$\textmu{}s, $\omega_{0}/2\pi=125$MHz,
$\omega_{1}/2\pi=50$kHz, and the amplitude of $n\left(t\right)/I_{0}$
is approximately $3.5$\%. At the output of the intensity modulator,
an optical splitter enables us to take approximatively 3\% of the
optical signal in order to measure the optical intensity in the fiber
by the mean of a second amplified photodiode. The resulting voltage
is digitized with a National Instrument PXI card at the sample rate
of 200 megasamples per second, or can be measured with a digital oscilloscope.

If we increase $I_{0}$ gradually, the intensity $I\left(t\right)$
undergoes a bifurcation diagram typical of nonlinear dynamical systems.
Figure \ref{fig:BifurcationDiagram} shows the excellent agreement
between the experimentally observed bifurcation diagram and the one
obtained by numerical simulations of the evolution equations (with
the bias $\varphi=0$). In this bifurcation diagram, the number of
bifurcations before reaching chaotic behavior is strongly affected
by the amount of noise in the system. Comparing this bifurcation diagram
to simulations is the most precise way we have to estimate the amount
of noise in our experimental setup. The estimated value is in agreement
with measurements carried out on each component of the setup separately.
We also verified that the thickness of the branches inside the bifurcation
diagram is mainly due to the noise level of oscilloscope.

\section*{Driven system}

We now consider the addition of an input term. The input is characterized
by a new time scale $T'\leq T$. The scalar input $u\left(n\right)$
(evolving in discrete time $n\in\mathbb{Z})$ and the input mask $b_{i}$
are transformed into a continuous input $s\left(t\right)$ as follows:
\begin{equation}
s\left(t\right)=b_{i}u\left(n\right)\quad\textrm{for}\quad t\in\left[nT'+\frac{iT'}{N},nT'+\frac{(i+1)T'}{N}\right],\quad i=0,\ldots,N-1\quad,\quad n\in\mathbb{Z}\label{eq:source}
\end{equation}
where $N$ is the number of nodes in the reservoir. The input mask
values $b_{i}$, $i=0,...,N-1$ are randomly chosen from a given distribution
(which may depend on the task). The time scale $\theta$ over which
the input $s\left(t\right)$ changes is: 
\begin{equation}
\theta=\frac{T'}{N}.\label{eq:NeuronLength}
\end{equation}

A voltage proportional to $s\left(t\right)$ is generated by a function
generator (at a sample rate of 200 Msamples/s) and added to the output
voltage of the amplified photodiode using a RF combiner placed at
the entrance of the RF amplifier. So the voltage $V\left(t\right)$
that drives the intensity modulator is a combination of the light
intensity and the input signal. The dynamical equations thus become

\begin{equation}
x(t)=\sin\left(\alpha\mathcal{F}_{\omega_{0}\omega_{1}}\left(x\left(t-T\right)+\frac{n\left(t\right)}{I_{0}}\right)+\beta\mathcal{F}_{\omega_{1}}\left(s\left(t\right)\right)+\varphi\right)\label{eq:xsinxSource}
\end{equation}
where $\beta$ is experimentally adjustable by varying the output
voltage amplitude of the function generator. In this equation, we
take into account that the RF combiner is placed before the RF amplifier,
and therefore that the source term is affected by the highpass filter
$\mathcal{F}_{\omega_{1}}$ of the RF amplifier. Because of the difference
between the time scales $T$ and $\omega_{1}/2\pi$, the effect of
the filter $\mathcal{F}_{\omega_{1}}$ is almost negligible. Its main
effect is to ensure that the effective source signal $\mathcal{F}_{\omega_{1}}\left(s\left(t\right)\right)$
has mean value zero.

In our experiments we take $T=8.5\mu s$. The number $N$ of variables
is taken in the range $50-200$. We take $T'=\frac{N}{N+1}T$ (see
figure \ref{fig:InputOutputAnd Discretization}). Thus for $N=50,$
we have $\theta=167ns$. The performance of the reservoir does not
depend on the exact value of $T'$ chosen, as long as $T'/T$ is not
a simple fraction (such as 1, 3/4, or 1/2), which would divide our
reservoir into different independent subsystems.

\section*{Discretized dynamics}

To obtain the discretized dynamics, we discretise the intensity along
the fiber according to
\begin{equation}
x_{i}(n)\simeq x\left(nT'+\left(i+\frac{1}{2}\right)\theta\right)\quad i=0,\ldots,N-1=T'/\theta\label{eq:xDiscretization-1}
\end{equation}
where we suppose that $\theta=T'/N=T/\left(N+k\right)$ with $k$
integer (see figure \ref{fig:InputOutputAnd Discretization}). We
thus have that the physical time $t$ is related to $n,i,k$ through
\begin{equation}
t=nT'+\left(i+\frac{1}{2}\right)\theta=\frac{\left(nN+\left(i+\frac{1}{2}\right)\right)T}{N+k}.\label{eq:ContinuousToDiscreteTime}
\end{equation}
Upon neglecting the effects of the filters $\mathcal{F}_{\omega_{0}}$
and $\mathcal{F}_{\omega_{1}}$ we obtain the discretized version
of eq. (\ref{eq:xsinxSource}). Note that the absence of synchronisation
($T'\neq T$, or equivalently $k>0$) completely modifies the dynamics
by coupling the discretised variables $x_{i}$ to each other. Note
also that the wrap arround effect (the second line in eq. (\ref{eq:discreteunsynchronizeddynamics}))
does not appear in traditional reservoir computing.

\section*{Numerical simulations}

Two different numerical models were developed to study the capabilities
of the network: a 'discretized' model, closer to the standard formulation
of Echo State Networks, and a 'continuous' model, which is as close
as possible to our experimental apparatus. 

In the 'discretized' version of the model we implement the discretization
described by eq. (\ref{eq:discreteunsynchronizeddynamics}). No noise
is considered, and the bandpass effects of the various components
are neglected. The sine nonlinearity and the topology of the network
are preserved. The optimal operating point of the system is found
by tuning the parameters $\alpha$ and $\beta$ in eq. (\ref{eq:discreteunsynchronizeddynamics}).
This model is used to set a performance goal for our experimental
system: if the performance of the experiment is close to the one of
the model, then our system is robust enough with respect to the noise
and the effects introduced by each of its components. Moreover, the
performance of the 'discretized' model is the same, within the experimental
error, to the one of traditional networks as reported in \cite{Rodan2010},
allowing us to validate the chosen nonlinearity and topology as good
choices for a reservoir computer.

The 'continuous' Matlab model we developed is instead as close as
possible to the experiment. All the signals in the simulation are
discretized at 200 Msamples/s. This corresponds to the sample rate
of the arbitrary waveform generator and the digitizer. All the components
are represented by their transfer function at their respective operating
point (sine function for the Mach-Zehnder modulator, responsivity
and transimpedance gains of the photodiodes, gain of the amplifier).
The collective frequency response of all the components is represented
by a bandpass filter with first order slopes. This is a reasonable
approximation to the exact slope of the bandpass filters which was
measured using a vector network analyzer in open loop configuration.
Noise is added to the signal at each noisy element of the system (dark
current of the photodiode, noise added by each amplifier...).

\begin{figure}
\includegraphics[scale=0.45]{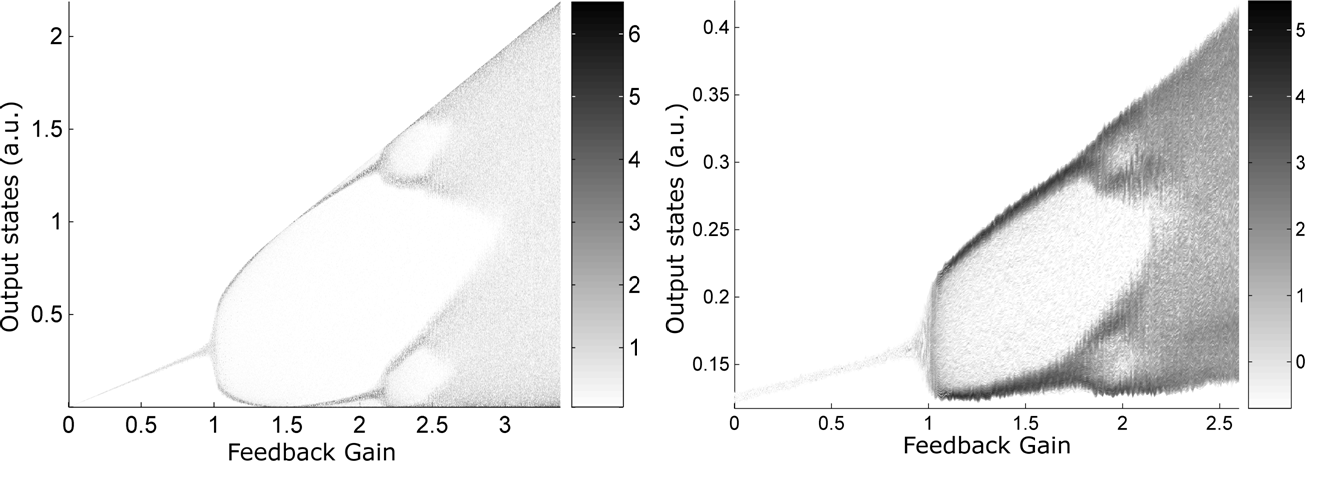}

\caption{\label{fig:BifurcationDiagram}Simulated (left panel) and measured
(right panel) bifurcation diagram. The vertical axis (output states)
is proportional to the optical intensity in the optical fiber for
the simulations, and to the voltage at the output of the readout photodiode
for the measurements (this voltage is equal to de optical intensity
multiplied by the gain of the amplified photodiode). The gray scale
represents the histogram of optical intensities inside the system
as a function of the feedback gain. When the feedback gain is lower
than unity, only one value of the light intensity is possible. For
feedback gain slightly larger than unity, the light intensity oscillates
between two values. For even larger feedback gain (around 2), the
nonbijective nature of the Mach-Zehnder modulator's transfer function
leads to oscillation between multiple light intensity levels or even
to a chaotic behavior. The number of bifurcations before reaching
the chaotic behavior is determined by the amount of noise inside the
system. The thickness of the branches in the measured bifurcation
diagram is due to the noise added by the oscilloscope.}
\end{figure}

The dynamics of the model correspond very closely to the experiment.
This is illustrated for instance by the clear agreement of the simulated
and measured bifurcation diagram (see fig. \ref{fig:BifurcationDiagram})
and by the clear agreement of the simulated and measured performances
on the channel equalization task, see Fig. \ref{fig:ChannelEqualization}
in the main text.

The continuous model allows us to easily explore the sensitivity to
parameters, such as noise level or the shape of the nonlinearity,
which can't always be reached with the experiment.

\section*{Post processing}

\begin{figure}
\includegraphics[scale=0.7]{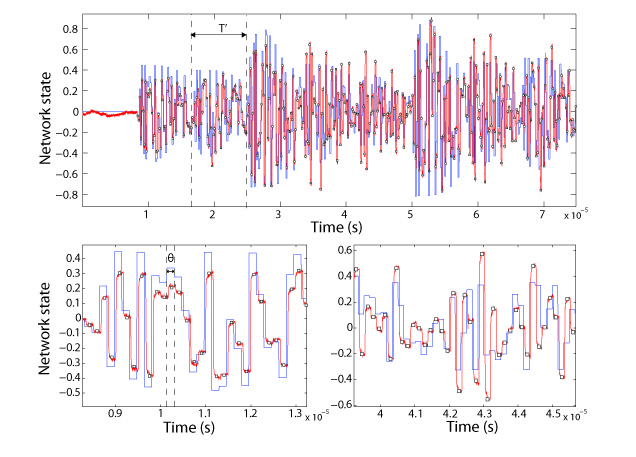}

\caption{\label{fig:InputOutputAnd Discretization} Operation of the experimental
reservoir. The blue line represents the masked input driving the reservoir;
the red line represents the reservoir response to the depicted input
as measured by the oscilloscope. Both have been normalized so that
their values lie between -1 and 1 for the whole length of the experiment.
The white squares along the red line represent the discretized values
of the reservoir states, obtained by averaging the reservoir output
on a time interval of amplitude $\theta/2$. The upper part of the
figure shows the reservoir operation for the first 8 inputs of the
reservoir; note that the input is zero up to $t\approx0.8\cdot10^{-5}$.
The bottom left panel shows a detail from the measurement on the response
to the very first input value, when no feedback is present yet, and
we only see the instantaneous response of the Mach-Zehnder. The panel
on the bottom right depicts the response of the system to the 5th
input, where the effect of the feedback is clearly visible: the amplitudes
of the reservoir states are no longer just related to the instantaneous
input, but are influenced by previous inputs as well.}
\end{figure}

The light intensity $I(t)$ in the fiber loop is converted to a voltage
by a photodiode and recorded by the digitizer operating at 200MS/s.
From the intensity $I(t)$ recorded during a time $T'$ we extract
$N$ discrete variable values $x_{i}(n)$, $i=0,...,N-1$. This is
carried out as follows. The intensity $I(t)$ is divided into $N$
pieces of duration $\theta$. We neglect the first quarter and the
last quarter of the data points recorded over the duration $\theta$
and associate to $x_{i}(n)$ the average of the remaining data points.
This procedure in which the beginning and end of each interval $\theta$
is omitted allows us to not be affected by the transients as the intensity
goes from one value to the other, and also allows an efficient synchronization
of our system. The estimator $\hat{y}(n)$ is then obtained by taking
a linear combination $\hat{y}(n)=\sum_{i}W_{i}x_{i}(n)$ where the
weights $W_{i}$ are optimized. This post processing is carried out
offline, on a computer. Fig. \ref{fig:InputOutputAnd Discretization}
shows an example of the input sent to the reservoir, the reservoir
response and the discretization operated on the reservoir output.

\section*{Tasks}

In our work we considered several tasks. We review them in detail.

\subsection*{Signal classification}

This is a simple task that we use for a first evaluation of the performance
of the reservoir. The input $u(n)$ to the system consists of random
sequences of sine and square waves discretized into $12$ points per
period. The mask values $b_{i}$ are drawn from the uniform distribution
over the interval $[0,+1]$. The reservoir size is taken to be $N=50$.
The output $\hat{y}(n)$ should be $1$ when the signal is the square
wave, and $0$ when it is the sine. The weights $W_{i}$ are obtained
by minimizing the MSE between $\hat{y}(n)$ and the ideal output.
Experimentally we obtain $\textrm{NMSE}\simeq1.5\ 10^{-3}$ , which
corresponds to essentially perfect operation for this task. These
experimental results are in close agreement with those obtained using
numerical simulations of our reservoir.

For comparison, practically the same task was studied previously in
\cite{Vandoorne2008}. An error rate (percentage of time the signal
was misclassified) of 2.5\% was obtained.

\subsection*{Nonlinear channel equalization}

The task is to reconstruct the input $d(n)\in\left\{ -3,-1,+1,+3\right\} $
of a noisy nonlinear wireless communication channel, given the output
$u(n)$ of the channel. The relation between $d(n)$ and $u(n)$ is
\begin{eqnarray}
q(n) & = & 0.08d(n+2)\text{\textendash}0.12d(n+1)+d(n)+0.18d(n\text{\textendash}1)\nonumber \\
 &  & \text{\textendash}0.1d(n\text{\textendash}2))+0.091d(n\text{\textendash}3)\text{\textendash}0.05d(n\text{\textendash}4)\label{eq:CheqSecondDefinition}\\
 &  & +0.04d(n\text{\textendash}5)+0.03d(n\text{\textendash}6)+0.01d(n\text{\textendash}7)\nonumber 
\end{eqnarray}

\begin{equation}
u(n)=q(n)+0.036q(n)^{2}\text{\textendash}0.011q(n)^{3}+\nu(n)\label{eq:CheqSecondNonlinNoise}
\end{equation}
 where $\nu(n)$ represents i.i.d. Gaussian noise with zero mean,
adjusted in power to yield signal-to-noise ratios ranging from 12
to 32 dB. This task was introduced in \cite{Mathews1994} and it was
shown in \cite{Jaeger2004} that reservoir computing could significantly
improve performance on this task.

In our study, the input mask is taken to be uniformly distributed
over the interval $[-1,+1]$. The reservoir size is taken to be $N=50$.
To perform the task, we first obtain an estimator $\hat{y}(n)$ of
the input by minimizing the MSE between $\hat{y}(n)$ and $d(n)$.
We then obtain an estimator $\hat{d}(n)$ by replacing $\hat{y}(n)$
by the discretized value $\left\{ -3,-1,+1,+3\right\} $ to which
it is closest. Finally we estimate the Symbol Error Rate (SER), i.e.,
the fraction of time that $\hat{d}(n)$ differs from $d(n)$. The
performance of the network is calculated as the average performance
over 10 different input sequences, in which the first 3000 samples
form the training set and the following 6000 samples form the test
set. The SER on the test set is then studied as a function of Signal
to Noise Ratio at the input. The values reported in figure \ref{fig:ChannelEqualization}
in the main text are the average SER for 10 different trials, while
the error bars represent the standard deviation of the SER for the
same trials. It should be noted that 6000 test steps are the maximum
number of steps that the arbitrary waveform generator in our setup
allows. Hence, when SERs approach $10^{-4}$, our average SERs include
trials where two errors, one error, or no error at all have been made
by the reservoir. This means that the error bars on the experimental
data where SERs are close to $10^{-4}$ might be overestimated. In
contrast, data from simulations do not suffer from this effect, as
we can arbitrarily increase the number of test samples for a more
precise measurement.

For comparison, at a SNR ratio of 28dB, the three models studied in
\cite{Mathews1994} gave SER of $2\cdot10^{-3}$,$4\cdot10^{-3}$,$1.5\cdot10^{-2}$,
while the reservoir studied in \cite{Jaeger2004} gave SERs of $1\cdot10^{-4}$
to $1\cdot10^{-5}$. At the same SNR of 28dB, our experimental system
gives a SER of $1.3\cdot10^{-4}$.

\subsection*{NARMA10}

In this task the aim is to reproduce the behavior of a nonlinear,
tenth-order system with random input drawn from a uniform distribution
over the interval $[0,0.5]$. The equation defining the target system
is given by

\begin{equation}
y(n+1)=0.3y(n)+0.05y(n)\left(\sum_{i=0}^{9}y(n-i)\right)+1.5u(n-9)u(n)+0.1\label{eq:NARMASecondDefinition}
\end{equation}

For this task, the mask is uniformly distributed over $[0,+1]$ and
the reservoir size is $N=50$. We first train our system using $1000$
time steps, then we test the system on a new sequence of $1000$ inputs.
The performance of the reservoir is measured by the Normalized Mean
Square Error of the estimator $\hat{y}(n)$, averaged over 10 different
pairs of train and test sequences. For this NARMA10 task, the best
performances ($\textrm{NMSE}=0.16$ for the discretized simulation,
$0.168$ for the continuous simulation and $0.167$ for the experiment
) are obtained in a very linear regime.

\subsection*{Isolated spoken digit recognition}

For the isolated spoken digit recognition task, the data is taken
from the NIST TI-46 corpus \cite{TexasInstruments}. It consists of
ten spoken digits (0...9), each one recorded ten times by five different
female speakers. These 500 spoken words are sampled at 12.5 kHz. This
spoken digit recording is preprocessed using the Lyon cochlear ear
model \cite{Lyon}. The input to the reservoir $u_{j}(n)$ consists
of an 86-dimensional state vector ($j=1,...,86$) with up to 130 time
steps. The number of variables is taken to be $N=200$. The input
mask is taken to be a $N\times86$ dimensional matrix $b_{ij}$ with
elements taken from the the set $\{-0.1,+0.1\}$ with equal probabilities.
The product $\sum_{j}b_{ij}u_{j}(n)$ of the mask with the input is
used to drive the reservoir. Ten linear classifiers $\hat{y}_{k}(n)$
($k=0,...,9$) are trained, each one associated to one digit. The
target function for $y_{k}(n)$ is +1 if the spoken digit is $k$,
and -1 otherwise. The classifiers are averaged in time, and a winner-takes-all
approach is applied to select the actual digit.

In our study, the 500 spoken words are divided in five subsets. We
trained the reservoir on four of the subsets, and then tested it on
the fifth one. This is repeated five times in order to use each subset
as test part. In this way we can test our system over all the speakers
and digits, and compute an average performance . The performance is
given in terms of the Word Error Rate, that is the fraction of digits
that are misclassified. We obtain a WER of $0.4\%$ which correspond
to 2 errors in 500 recognized digits.

For comparison, in \cite{Verstraeten2005}, where Reservoir Computing
was first used on this spoken digit benchmark, a WER of 4.3\% was
reported for a reservoir of size 1232. In \cite{Verstraeten2006}
a WER of 0.2\% was obtained for a reservoir of size 308 and using
the winner-takes-all approach. In \cite{Rodan2011} a WER of 1.3\%
for reservoirs of size 200 is reported. The experimental reservoir
reported in \cite{Appeltant} gives a WER of 0.2\% using a reservoir
of size 400. The Sphinx-4 system \cite{Walker2004} uses a completely
different method based on Hidden Markov Models and achieves a WER
of 0.55\% on the same data set. 

\pagebreak{}


\begin{thebibliography}{10}
\bibitem{Caulfield2010} H.~John Caulfield and Shlomi Dolev. \newblock {Why future supercomputing requires optics}. \newblock {\em Nature Photonics}, 4(5):261--263, 2010.
\bibitem{Jaeger2001} H.~Jaeger. \newblock {The "echo state" approach to analysing and training recurrent neural   networks}. \newblock Technical report, Technical Report GMD Report 148, German National   Research Center for Information Technology, 2001.
\bibitem{Jaeger2001a} H.~Jaeger. \newblock {Short term memory in echo state networks}. \newblock Technical report, Technical Report GMD Report 152, German National   Research Center for Information Technology, 2001.
\bibitem{Jaeger2004} Herbert Jaeger and Harald Haas. \newblock {Harnessing nonlinearity: predicting chaotic systems and saving   energy in wireless communication.} \newblock {\em Science (New York, N.Y.)}, 304(5667):78--80, 2004.
\bibitem{Legenstein2005} Robert Legenstein and Wolfgang Maass. \newblock {\em New Directions in Statistical Signal Processing: From Systems to   Brain}, chapter {What makes a dynamical system computationally powerful?},   pages 127--154. \newblock MIT Press, 2005.
\bibitem{Maass2002} Wolfgang Maass, T.~Natschlager, and Henry Markram. \newblock {Real-time computing without stable states: A new framework for   neural computation based on perturbations}. \newblock {\em Neural computation}, 14(11):2531--2560, 2002.
\bibitem{Steil} J.J. Steil. \newblock {Backpropagation-decorrelation: online recurrent learning with O(N)   complexity}. \newblock {\em 2004 IEEE International Joint Conference on Neural Networks   (IEEE Cat. No.04CH37541)}, pages 843--848.
\bibitem{Verstraeten2007} David Verstraeten, Benjamin Schrauwen, M~D'Haene, and Dirk Stroobandt. \newblock {An experimental unification of reservoir computing methods.} \newblock {\em Neural networks : the official journal of the International   Neural Network Society}, 20(3):391--403, 2007.
\bibitem{Lukosevicius2009} Mantas Luko\v{s}evi\v{c}ius and Herbert Jaeger. \newblock {Reservoir computing approaches to recurrent neural network   training}. \newblock {\em Computer Science Review}, 3(3):127--149, 2009.
\bibitem{Hammer2009} Barbara Hammer and Benjamin Schrauwen. \newblock {Recent advances in efficient learning of recurrent networks}. \newblock In {\em Proceedings of the European Symposium on Artificial Neural   Networks}, pages 213--216, 2009.
\bibitem{NeuralForecastingCompetition} {http://www.neural-forecasting-competition.com/NN3/index.htm}.
\bibitem{Verstraeten2006} David Verstraeten, Benjamin Schrauwen, and Dirk Stroobandt. \newblock {Reservoir-based techniques for speech recognition}. \newblock In {\em The 2006 IEEE International Joint Conference on Neural   Network Proceedings}, pages 1050--1053. IEEE, 2006.
\bibitem{Triefenbach2010} Fabian Triefenbach, A.~Jalalvand, Benjamin Schrauwen, and J.~Martens. \newblock {Phoneme recognition with large hierarchical reservoirs}. \newblock {\em Advances in Neural Information Processing Systems}, 23:1--9,   2010.
\bibitem{Jaeger2007} Herbert Jaeger, Mantas Lukosevicius, Dan Popovici, and Udo Siewert. \newblock {Optimization and applications of echo state networks with   leaky-integrator neurons.} \newblock {\em Neural networks : the official journal of the International   Neural Network Society}, 20(3):335--52, 2007.
\bibitem{Fernando2003} Chrisantha Fernando and Sampsa Sojakka. \newblock Pattern recognition in a bucket. \newblock In Wolfgang Banzhaf, Jens Ziegler, Thomas Christaller, Peter   Dittrich, and Jan Kim, editors, {\em Advances in Artificial Life}, volume   2801 of {\em Lecture Notes in Computer Science}, pages 588--597. Springer   Berlin / Heidelberg, 2003.
\bibitem{Schurmann2004} Felix Sch�rmann, Karlheinz Meier, and Johannes Schemmel. \newblock Edge of chaos computation in mixed-mode vlsi - a hard liquid. \newblock In {\em Advances in Neural Information Processing Systems}. MIT   Press, 2005.
\bibitem{Vandoorne2008} Kristof Vandoorne, Wouter Dierckx, Benjamin Schrauwen, David Verstraeten, Roel   Baets, Peter Bienstman, and Jan {Van Campenhout}. \newblock {Toward optical signal processing using photonic reservoir   computing.} \newblock {\em Optics express}, 16(15):11182--92, 2008.  
\bibitem{Appeltant} L.~Appeltant, M.C. Soriano, G.~{Van der Sande}, J.~Danckaert, Serge Massar,   J.~Dambre, Benjamin Schrauwen, C.~R. Mirasso, and I.~Fischer. \newblock {Information processing using a single dynamical node as complex   system}. \newblock {\em submitted to Nat. Comm.}, 2011.
\bibitem{Rodan2011} Ali Rodan and Peter Tino. \newblock {Minimum complexity echo state network.} \newblock {\em IEEE transactions on neural networks}, 22(1):131--44, 2011.
\bibitem{Erneux} Thomas Erneux. \newblock {\em Applied Delay Differential Equations}. \newblock Springer, 2009.
\bibitem{Larger2005} Laurent Larger, Pierre-Ambroise Lacourt, St\'{e}phane Poinsot, and Marc Hanna. \newblock {From flow to map in an experimental high-dimensional electro-optic   nonlinear delay oscillator}. \newblock {\em Physical Review Letters}, 95(4):1--4, 2005.
\bibitem{ChemboKouomou2005} Yanne~K. Chembo, Pere Colet, Laurent Larger, and Nicolas Gastaud. \newblock {Chaotic Breathers in Delayed Electro-Optical Systems}. \newblock {\em Physical Review Letters}, 95(20):2--5, 2005.
\bibitem{Peil2009} Michael Peil, Maxime Jacquot, Yanne~K. Chembo, Laurent Larger, and Thomas   Erneux. \newblock {Routes to chaos and multiple time scale dynamics in broadband   bandpass nonlinear delay electro-optic oscillators}. \newblock {\em Physical Review E}, 79(2):1--15, 2009.
\bibitem{Rodan2010} Ali Rodan and Peter Tino. \newblock {Simple Deterministically Constructed Recurrent Neural Networks}. \newblock {\em Intelligent Data Engineering and Automated Learning - IDEAL   2010}, pages 267--274, 2010.
\bibitem{Mathews1994} V.~John Mathews and Junghsi Lee. \newblock {Adaptive algorithms for bilinear filtering}. \newblock {\em Proceedings of SPIE}, 2296(1):317--327, 1994.
\bibitem{Verstraeten2005} David Verstraeten, Benjamin Schrauwen, and Dirk Stroobandt. \newblock {Isolated word recognition using a liquid state machine}. \newblock In {\em Proceedings of the 13th European Symposium on Artificial   Neural Networks (ESANN)}, pages 435--440, 2005.
\bibitem{Walker2004} Willie Walker, Paul Lamere, Philip Kwok, Bhiksha Raj, Rita Singh, Evandro   Gouvea, Peter Wolf, and Joe Woelfel. \newblock Sphinx-4: a flexible open source framework for speech recognition. \newblock Technical report, Mountain View, CA, USA, 2004.
\bibitem{Atiya2000} Amir~F. Atiya and Alexander~G. Parlos. \newblock {New results on recurrent network training: unifying the algorithms   and accelerating convergence.} \newblock {\em IEEE transactions on neural networks}, 11(3):697--709, 2000.
\bibitem{Jaeger2002} H.~Jaeger. \newblock {Adaptive Nonlinear System Identification with Echo State Networks}. \newblock In {\em Advances in Neural Information Processing Systems}, volume~8,   pages 593--600. MIT Press, 2002.
\bibitem{TexasInstruments} {Texas Instruments-Developed 46-Word Speaker-Dependent Isolated Word Corpus   (TI46), September 1991, NIST Speech Disc 7-1.1 (1 disc)}, 1991.
\bibitem{Lyon} R.~Lyon. \newblock {A computational model of filtering, detection, and compression in   the cochlea}. \newblock In {\em ICASSP '82. IEEE International Conference on Acoustics,   Speech, and Signal Processing}, pages 1282--1285. Institute of Electrical and   Electronics Engineers, 1982.
\bibitem{Ananthanarayanan2009} Rajagopal Ananthanarayanan, Steven~K. Esser, Horst~D. Simon, and Dharmendra~S.   Modha. \newblock {\em {The cat is out of the bag: : cortical simulations with 1\^{}09   neurons, 10\^{}13 synapses}}. \newblock SC '09. ACM, New York, NY, USA, 2009.
\bibitem{Verstraeten} David Verstraeten, Joni Dambre, Benjamin Schrauwen, and Serge Massar. \newblock Linear and nonlinear memory capacity of dynamical systems. \newblock {\em in preparation}.
\bibitem{Verstraeten2010} David Verstraeten, Joni Dambre, Xavier Dutoit, and Benjamin Schrauwen. \newblock {Memory versus non-linearity in reservoirs}. \newblock In {\em The 2010 International Joint Conference on Neural Networks   (IJCNN)}, pages 1--8. IEEE, July 2010.
\bibitem{Schrauwen2008} Benjamin Schrauwen, M~Wardermann, David Verstraeten, J.J. Steil, and Dirk   Stroobandt. \newblock {Improving reservoirs using intrinsic plasticity}. \newblock {\em Neurocomputing}, 71(7-9):1159--1171, 2008.
\bibitem{Nikolic} Danko Nikolic, Stefan H�usler, Wolf Singer, and Wolfgang Maass. \newblock Temporal dynamics of information content carried by neurons in the   primary visual cortex. \newblock {\em Advances in Neural Information Processing Systems},   19:1041--1048, 2007.
\bibitem{Hermans2010} Michiel Hermans and Benjamin Schrauwen. \newblock {Memory in linear recurrent neural networks in continuous time.} \newblock {\em Neural networks : the official journal of the International   Neural Network Society}, 23(3):341--55, 2010.
\end{thebibliography}
\end{document}